\documentclass[twocolumn,showpacs,showkeys,superscriptaddress]{revtex4}
\usepackage{graphicx}
\usepackage{bm}
\usepackage{color}
\usepackage{amsmath}
\usepackage{natbib}

\begin{document}

\title{Analogue black hole in magnetohydrodynamics}

\author{Felipe A. Asenjo}
\email{fasenjo@levlan.ciencias.uchile.cl}
\affiliation{Departamento de F\'\i sica, Facultad de Ciencias, Universidad de Chile, Casilla 653, Santiago, Chile.}
\affiliation{Departamento de Ciencias, Facultad de Artes Liberales,  Universidad Adolfo Ib\'a\~nez, Diagonal Las Torres 2640, Pe\~nalol\'en, Santiago, Chile.}

\author{Nelson Zamorano}
\email{nzamora@dfi.uchile.cl}
\affiliation{Departamento de F\'\i sica, Facultad de Ciencias F\'isicas y Matem\'aticas, Universidad de Chile, Casilla 487-3, Santiago, Chile.}

\date{\today}

\begin{abstract}
We consider an irrotational plasma fluid evolving under the effect of a background magnetic field. The magnetohydrodynamic formalism is used to describe the electromagnetic waves and the dynamics is described by a scalar field that follows a second order differential equation. This equation can also be recovered as the wave equation associated to a field in a curved space-time. Through this analogy we recreate a sonic horizon, equivalent to those found in perfect fluid theories. However, in this case, the magnetic field creates a pressure in the plasma which contributes to the magnetoacoustic speed that builds the horizon.  This effect enhances the temperature produced by the Hawking radiation expected from this analogue black hole, and eventually, making its experimental detection worth to consider.

\end{abstract}


\maketitle


\section{Introduction}

Thirty years after Unruh's seminal paper \cite{unruh1}, the study of the field of artificial black holes is attracting the interest of a large number of researchers. The result found by Unruh emerges from the similarity between fluid theory and general relativity at the kinematical level. Those systems whose dynamics is controlled by a second order differential equations display the same features than  a dynamical field evolving  on a curved space-time. In these cases, an effective metric can be extracted from the equation of motion of the fluid, and this metric display similar properties to those found in the neighborhood of a black hole event horizon.

A  black hole exhibits what is called the Hawking radiation which is associated to the event horizon with a temperature proportional to the surface gravity of the black hole \cite{hawking}. However, the detection of this radiation is a real challenge. In this context, if the analog black holes may reproduce all the features of the Hawking radiation, it offers the opportunity to search for the quantum instability proposed by Hawking. For this reason, there are several ongoing efforts searching for this radiation. The experimental verification and the possibility of studying their properties within this new context, would show the robustness of the Hawking effect.

Some of the settings for analog black holes include, perfect fluids, where Unruh originally found a horizon generated by the relative changes between the acoustic and the flow speeds \cite{unruh1,unruh2}. This effect  has also been studied within the context of gravity waves \cite{gravit} and dielectric mediums \cite{dielec}, looking for systems where a higher Hawking temperature could be measured. The search has been concentrated on the Bose-Einstein condensates \cite{garay,visser,giova,garay2}, where it has been claimed a measured evidence of the Hawking radiation \cite{fagno2}. In this context, they estimated a spectrum of phonons at $\sim7\cdot 10^{-8}$K \cite{visser}.
The use of waveguides have also been proposed to detect the Hawking radiation since the electromagnetic radiation can be controlled and detected much easier than sound \cite{ralf}. Using this mechanism, the experimental formation of an artificial horizon have been achieved in optical fibers \cite{philbin}.

Recently, an ultrashort  laser pulsed experiment has been performed to verify the Hawking radiation \cite{belgiorno}. The results are encouraging even though they have rose some controversy concerning the thermal spectrum of the radiation measured \cite{comments}. Another experiment using an open channel fluid with an obstacle interposed \cite{unruh3} has allowed to mimic a white hole, verifying the thermal nature of the emitted radiation, as underlined by the authors.

Here we address the occurrence of analogue black holes in plasmas. We compare the Hawking effect associated to sound waves in this context with the known ones using neutral fluids.
Magnetized plasmas offer a new kind of contribution to the analogue black holes, since they are the simplest system where a fluid of charged particles interact with a magnetic field.

Concretely, we study a magnetized plasma using the ideal magnetohydrodynamics (MHD) theory. In this approach, a plasma, composed by electrons and ions, is described as a single fluid. The equations for MHD are constructed as a linear combination of the fluid variables of the two kind of charged particles \cite{shercliff}.

For a quasineutral plasma composed by electrons and ions, a one-fluid mass density can be defined as $\rho\,=\,n\,(M+m)$, where $n$ is the density number, equal for electron and ion fluid, and $M$ and $m$ are the ion and electron masses respectively ($M\gg m$). Similarly, a single fluid velocity  is defined as ${\bf v}=(M {\bf v}_i+m{\bf v}_e)/(M+m)$, where ${\bf v}_i$ and ${\bf v}_e$ are the velocities of the ion and electron fluid, respectively.

When the Hall current effect is neglected and a null resistivity plasma is considered, it is possible to find a simple set of MHD equations, called ideal MHD. In this case, the Ohm law is written in the simple form ${\bf E}+{\bf v}\times{\bf B}=0$, and the displacement current $\partial{\bf E}/\partial t$ is neglected from the Ampere's law.

The equations that enter in this description for the ideal MHD theory are, the continuity equation
\begin{equation}
 \frac{\partial \rho}{\partial t}+\nabla\cdot\left(\rho {\bf v}\right)=0\, ,
\label{eccont}
\end{equation}
the equations of motion for a single fluid
\begin{equation}
 \rho\left(\frac{\partial}{\partial t}+{\bf v}\cdot\nabla\right){\bf v}=-\nabla\left(p+\frac{1}{8\pi}|{\bf B}|^2\right)+\frac{1}{4\pi}\left({\bf B}\cdot\nabla\right){\bf B}\, ,
\label{ecmom}\end{equation}
where $p$ is the pressure of the single fluid \cite{comm}, and  Maxwell�s equations are
\begin{equation}
 \frac{\partial {\bf B}}{\partial t}=\left({\bf B}\cdot\nabla\right){\bf v}-{\bf B}\left(\nabla\cdot{\bf v}\right)-\left({\bf v}\cdot\nabla\right){\bf B}\, .
\label{ecMax}\end{equation}

Eqs. \eqref{ecmom} and \eqref{ecMax} are obtained using the current density ${\bf J}=(c/4\pi)\nabla\times{\bf B}$, from Ampere's law, and the Ohm law respectively.

We can use the ideal MHD equations to find the propagation of waves in any configuration of electromagnetic and fluid fields.

\section{Magnetohydrodynamical black hole}
\label{sec2}

Since we follow a perturbation scheme, we define a subindex $0$ for the non-perturbed background fields, and a subindex $1$ for the first order perturbations. So, $\rho\,=\,\rho_0\,+\,\rho_1$, ${\bf v}={\bf v}_0+{\bf v}_1$, ${\bf B}={\bf B}_0+{\bf B}_1$ and $p\,=\,p_0\,+\,p_1$. The background magnetic field ${\bf B}_0$ and the background density $\rho_0$ remain  constants in this context, and consequently the background becomes incompressible, $\nabla\cdot{\bf v}_0=0$. 

In general, the magnetic field introduces  a vorticity into the plasma. However, the longitudinal modes accept an irrotational solution for the plasma velocity. The magnetoacoustic mode is one of these longitudinal modes where the wave is longitudinal in the perturbation of the fluid velocity, although the perturbation of the magnetic field is transverse. Considering perturbations of the form $\exp(i{\bf k}\cdot {\bf x}-i\omega t)$ in the perturbed quantities, where ${\bf k}$ and $\omega$ are the wavenumber and its frequency respectively, the propagation is transversal to the perturbed field ${\bf B}_1$, i.e. ${\bf B}_1\cdot{\bf k}=0$. This condition is relevant for the propagation of magnetoacoustic modes.

We also impose the following constrains: the perturbation ${\bf B}_1$ is chosen to be parallel to the background magnetic field ${\bf B}_0\times{\bf B}_1=0$, and then ${\bf B}_0\cdot{\bf k}\,=\,0$. This implies that $({\bf B}_0\cdot\nabla){\bf
 B}_1=0$. Besides, we choose ${\bf B}_0\cdot{\bf v}_0=0$ for the longitudinal mode. Given these assumptions, we proceed to write Eqs.\eqref{eccont}-\eqref{ecMax} keeping only first order terms
\begin{equation}
 \left(\frac{\partial}{\partial t}+{\bf v}_0\cdot\nabla\right)\rho_1+\rho_0 \nabla\cdot{\bf v}_1=0\, ,
\label{eccont1}
\end{equation}
\begin{equation}
 \rho_0\frac{\partial{\bf v}_1}{\partial t}+\rho_0\nabla\left({\bf v}_0\cdot{\bf v}_1\right)=-\nabla p_1-\frac{1}{4\pi}\nabla\left({\bf B}_0\cdot{\bf B}_1\right)\, ,
\label{ecmom1}\end{equation}
\begin{equation}
\left(\frac{\partial}{\partial t}+{\bf v}_0\cdot\nabla\right){\bf B}_1=\left({\bf B}_1\cdot\nabla\right){\bf v}_0+\left({\bf B}_0\cdot\nabla\right){\bf v}_1-{\bf B}_0\left(\nabla\cdot{\bf v}_1\right)\, ,
\label{ecMax1}\end{equation}
where the perturbed velocity is longitudinal. The Ohm's law is satisfied to first order: ${\bf E}_1\,+\,{\bf v}_1\times{\bf B}_0\,+\,{\bf v}_0\times{\bf B}_1$\,=\,0.

From Eq.\eqref{ecmom1} we have that $\nabla\times{\bf v}_1\,=\,0$, the perturbed velocity aligns itself in the direction of the wave propagation ${\bf k}$ and, as a consequence, the plasma fluid remains irrotational.
This is the main step behind our approach. When a fluid is irrotational, its velocity can be described by a single scalar field potential $\psi_1$. The first order perturbation velocity is then, determined by this potential:  ${\bf v}_1\,=\,\nabla\,\psi_1$.

For the plasma to remain irrotational as it evolves in time, the pressure associated to the fluid should depend only on the density \cite{novello}. In our case, we have chosen an isothermal plasma, such that $\nabla\, p_1\,=\,v_s^2\, \nabla\rho_1$, where $v_s\,\equiv\,\left(\,\gamma p_0/\rho_0\,\right)^{1/2}$ is the adiabatic sound speed. When ions and electrons are involved in a fluid, the sound speed is given by $v_s^2\,=\,k_B\, (\,\gamma_i\, T_i\,+\,\gamma_e\,T_e\,)/(M+m)\,\approx\, k_B\,(\gamma_i T_i\,+\,\gamma_e\,T_e)/M$, where $T_i$  and $T_e$ are the ion and electron fluid temperatures respectively and $k_B$ is the Boltzmann constant  \cite{chen}. In this case, we use $\gamma=5/3$.

With this choice for the magnetic fields and with the condition of an irrotational plasma, we have that ${\bf B}_0\,\cdot\,{\bf v}_1\,=\,0$ . Then, from equations \eqref{eccont1} and Eqs.\eqref{ecMax1}, we get
\begin{equation}
 \left(\frac{\partial}{\partial t}+{\bf v}_0\cdot\nabla\right)\left({\bf B}_0\cdot{\bf B}_1-\xi_1 B_0^2\right)=0\, ,
\label{ecMax2}\end{equation}
where $\xi_1=\rho_1/\rho_0$ and $B_0^2={\bf B}_0\cdot{\bf B}_0$.
Therefore, using \eqref{ecMax2} and neglecting the integration constant, the Eqs. \eqref{eccont1} and \eqref{ecmom1} become
\begin{equation}
 \left(\frac{\partial}{\partial t}+{\bf v}_0\cdot\nabla\right)\xi_1+ \nabla^2 \psi_1=0\, ,
\label{eccont2}
\end{equation}
\begin{equation}
\left(\frac{\partial}{\partial t}+{\bf v}_0\cdot\nabla\right){\psi_1}=-c^2 \xi_1\, ,
\label{ecmom2}\end{equation}
where $c$ is the magnetoacoustic speed \cite{priest} defined as
\begin{equation}
c=\left(v_s^2+v_A^2\right)^{1/2}\, ,
\label{velocidfluidAlfven}
\end{equation}
and
\begin{equation}
v_A\,=\,\frac{B_0}{\sqrt{4\pi\rho_0}}\, ,
\label{Alfven}
\end{equation}
 is the Alfv\'en velocity.

The equations \eqref{eccont2} and \eqref{ecmom2} for $\xi_1$ and $\psi_1$ show the mutual influence -as expected- from the dynamics of the density and the velocity perturbations for the irrotational plasma fluid. The magnetic field is proportional to the Alfv\'en velocity and, through this velocity determines the magnetoacoustic speed, see \eqref{velocidfluidAlfven} and \eqref{Alfven}. The linearization of both equations gives the propagation of the magnetoacoustic modes in the MHD formalism \cite{priest}. This mode propagates perpendicularly to the magnetic field compressing and releasing both the lines of force and the conducting fluid.

 Eqs. \eqref{eccont2} and \eqref{ecmom2} can be combined into a single second-order differential equation for the scalar field $\psi_1$
\begin{eqnarray}
\frac{\partial}{\partial t}\left(\rho_0\frac{\partial\psi_1}{\partial t}+\rho_0{\bf v}_0\cdot\nabla\psi_1\right)&+&\rho_0{\bf v}_0\cdot\nabla\left(\frac{\partial\psi_1}{\partial t}+{\bf v}_0\cdot\nabla\psi_1\right)\nonumber\\
&&-\rho_0c^2\nabla^2\psi_1=0\, ,
\label{Klienpsi}\end{eqnarray}
where the coefficients are given by the background quantities of the plasma. This equation contains the same information about the propagation of the magnetoacoustic waves than the two previous equations.

 Eq. \eqref{Klienpsi} can be re-arranged to appear as the Klein-Gordon equation for a massless scalar field in a curved space-time: $\partial_\mu\left(\sqrt{-g} g^{\mu\nu}\partial_\nu\psi_1\right)/\sqrt{-g}=0$. To achieve this, we need to introduce the effective Unruh metric \cite{unruh1} (with the Greek indices running from 0-3)
\begin{equation}
g^{\mu\nu}=\frac{1}{\rho_0 c}\left(\begin{array}{cc}
 -1 & -v_0^i \\
-v_0^j & -v_0^iv_0^j+c^2\delta^{ij}
\end{array}\right)\, ,
\end{equation}
where $v_0^j$ is the $j$ component of ${\bf v}_0$ (the Latin indices run from 1-3), and $g\,\equiv\,[\det g^{\mu\nu}]^{-1}=-c^2$. The inverse of this metric is
\begin{equation}
g_{\mu\nu}=\frac{\rho_0}{c}\left(\begin{array}{cc}
 v_0^2-c^2 & -v_0^j \\
-v_0^i & \delta^{ij}
\end{array}\right)\, ,
\label{metric}
\end{equation}
where $v_0^2={\bf v}_0\cdot{\bf v}_0$.

The metric \eqref{metric} can be interpreted as an effective curved space-time where these waves propagate. The interval extracted from this geometry is
\begin{eqnarray}
ds^2&=&g_{\mu\nu}dx^\mu dx^\nu\nonumber\\
&=&\frac{\rho_0}{c}\left(\left[v_0^2-c^2\right]dt^2-2 dt {\bf v}_0\cdot d{\bf x} +d{\bf x}\cdot d{\bf x}\right)\, .
\label{interval}\end{eqnarray}
Assuming that the background flow is spherical symmetric and stationary, and defining the new time $d\tau=dt-v_0 dr/(v_0^2-c^2)$ \cite{unruh1}, we can write the interval \eqref{interval} as
\begin{equation}
ds^2=\frac{\rho_0}{c}\left(\left(v_0^2-c^2\right)d\tau^2-\frac{c^2 dr^2}{v_0^2-c^2}-r^2d\Omega^2\right)\, ,
\label{interval2}
\end{equation}
where $d\Omega^2=d\theta^2+\sin^2\theta d\phi^2$. In general, $v_0$ depends on the radius.

The metric \eqref{interval2} assuming a velocity $v_0\,\approx\,-c\,+\,\alpha\,(r-R)\,+...$ displays properties similar to a Schwarzschild black hole as first proved by Unruh \cite{unruh1}. Black holes are characterized by an event horizon, a mathematical null surface that splits the trajectory of light rays
depending if they travel inside or outside this surface. This is a kinematical consequence of the event horizon existence. On the other hand, the entropy as proportional to the area of a black hole is a dynamical consequence of Einstein equations \cite{visser1998}.
In the analogue black hole, the event horizon is a kinematic effect and is characterized for the velocity of the fluid reaching the sonic velocity. It is explained in more detail with a drawing in the next section. The temperature arises in the analog black hole from the quantizing of the scalar field characterizing the velocity perturbation. This is the main result from the first paper by Unruh \cite{unruh1}. The procedure is generic and it is possible to follow step by step for this case. We return to this subject in the following section.

\section{Hawking radiation in plasmas}

The possibility of the existence, in the sense of measurable, of a thermal radiation coming from an acoustic horizon caused by a supersonic flow, will set the thermal radiation generated on the event horizon of a black hole in a different perspective. For this reason we examine the Hawking temperature obtained for a MHD plasma within this context.

In Unruh's approach \cite{unruh1} there exists a scalar field $\psi$, as a perturbation around some background solution $\psi_0$ related to the fluid velocity potential. Expanding this field $\psi$ in terms of creation and annihilation operators and after some canonical manipulation he concluded that for this fluid with a sonic horizon, there exists a thermal spectrum of sound waves radiated. The expression obtained for the temperature of the analogue black hole can be compared term by term with the Hawking expression and given a corresponding physical interpretation.

Following the same steps applied to the MHD fluid, it occurs that near the acoustic horizon, the analogue black hole will emit magnetoacoustic waves with a thermal spectrum with a temperature  given by
\begin{equation}
T_H=\frac{\hbar g_H}{2\pi\, k_B\, c}\, ,
\end{equation}
where $g_H$ is the \textit{surface gravity} on the acoustic horizon \cite{novello} given by the following expression
\begin{equation}
g_H=c\left|\frac{\partial}{\partial n}(c-v_0) \right|\, ,
\end{equation}
and $\partial/\partial n$ represents the derivative of the plasma fluid velocity taken normal to the cross section of the horizon.
In our case, $c\,>\,v_s$ always due to the expression of the Alfv\'en velocity. Thus, the background magnetic field, within this context, increases the value of the Hawking temperature compared with the effect of a neutral fluid.

\begin{figure}[h!]
\includegraphics[height=4cm]{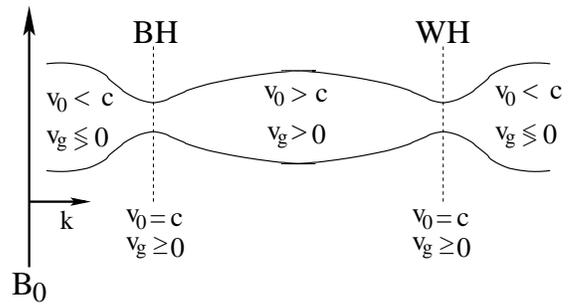}
\caption{\label{figura2} Laval nozzle and background magnetic field. First, the wave is subsonic with $v_g<0$ or $v_g>0$. The narrowest part of the nozzle represents the horizon with $v_0=c$ and $v_g\geq 0$ (the black hole, BH). Then, the waves become supersonic with $v_g>0$ only. Finally, they reach the other horizon (the white hole, WH) to return to be subsonic.}
\end{figure}

We propose the following setting to achieve a magnetoacoustic wave propagating through a background magnetic field and its thermal radiation associated, as described in Sec. \ref{sec2}. Let's consider a plasma traveling in a cylindrical configuration under the influence of a background magnetic field in the transversal direction ${\bf B}_0=B_0 \hat z$ (see Fig.\ref{figura2}). These waves will propagate in the direction of increasing $\theta$. We need to introduce an approximation at this point, the radius of this cylindrical configuration must be large compared with the size of the nozzle. In this way when we compute the relevant physics in this region assuming a straight motion of the plasma.

Following Ref.\cite{novello}, we set a Laval nozzle in some region of this cylindrical device. Within this device, when the cross section shrinks, reaching its minimal section, the plasma flowing through it, will go supersonic. For this condition to occur, as mentioned above, the radius of the cylindrical device must be larger than the length characterizing the nozzle. In other words, the Larmor radius of the plasma constituents induced by the background and perturbed magnetic fields must be much larger than the lenght of the Laval nozzle. Within this approximation the plasma can be considered as a longitudinal flow in that region. In Fig.\ref{figura2} we show a diagram of the Laval nozzle and the background magnetic field.

If the space-time dependence of this wave is $\exp(i {\bf k}\cdot{\bf r}-i\omega t)$, where ${\bf k}=k\hat\theta$, then from Eqs. \eqref{eccont2} and \eqref{ecmom2}, we can obtain the dispersion relation for the magnetoacoustic waves $v_g\,=\,v_0\,\pm\, c$  \cite{priest}, where $v_g\,=\,\partial \omega/\partial k$ is the group velocity and the drift velocity of the plasma lies mainly on the longitudinal direction ${\bf v}_0\,=\,v_0\hat{\theta}$. It is easy to see that if the fluid is subsonic, then $v_g\,<\,0$ for the reflected waves in the fluid and $v_g\,>\,0$ for those transmitted, therefore the waves travel forward or backward along the $\hat{\theta}$ direction under this condition. At the horizon, the center of the nozzle, when the velocity becomes supersonic $v_0\,=\,c$, occurs that $v_g\,\geq\, 0$ for all waves, and every wave crosses the horizon without return. This nozzle behaves as a black hole event horizon. At the other side of the nozzle, those waves become supersonic with a group velocity $v_g\,>\,0$ and  $v_g\,>\,c$ always, and they keep traveling forward (to the right).

A time inverse situation occurs in the following nozzle, as the velocity decreases, reaching $v_0\,=\,c$ at the nozzle and there, the velocities are equal or larger than the sound speed and all the fluid moves to the right. It is the equivalent of a white hole.

The Laval nozzle is the key to generate a supersonic plasma fluid. It plays the role of the acoustic horizon.
If the radius of the narrowest part of the nozzle is $R$, we can approximate
\begin{equation}
\frac{1}{c}\left.\frac{\partial(c-v_0)}{\partial n}\right|_H\approx\frac{1}{R}\, .
\label{approx111}
\end{equation}

This approximation is valid because the magnetoacoustic speed is constant. Then, the surface gravity will be $g_H\,\approx\, c^2/R$. This is essentially the same result obtained for the surface gravity $g_H\,\approx \,c^2/\sqrt{2\, A_H}$ considering the cross section area $A_H\sim R^2$ of the Laval nozzle \cite{visser}.

Thus, applying the same formalism used in previous works \cite{unruh1,visser}, the corresponding  Hawking temperature is
\begin{equation}
T_H\approx\frac{\hbar c}{2\pi k_B R}\, .
\end{equation}

This is the main result of this work. Also, when we compare this temperature with previous cases, it appears that we always can achieve a larger temperature compared with a neutral fluid because $c\,>\,v_s$. An interesting case occurs when the Alfv\'en velocity is much larger than the sound speed. The temperature of a tenuous laboratory plasma is around $10^4~$K, and therefore the sound speed will be of the order of $10^6~\mbox{cm}/\mbox{s}$. On the other hand, for typical values of the number density of $n_0=10^{12}$ cm$^{-3}$ and a magnetic field of $B_0\,=\,10^4$ gauss, the Alfv\'en velocity is of the order of $10^{9}~\mbox{cm}/\mbox{s}$, and hence $v_s\ll v_A$. In this case, the magnetic pressure becomes larger than the fluid pressure $B_0^2/4\,\pi \gg  p_0$, and then $c\approx v_A$. The Hawking temperature now will depend mainly on the background magnetic field through the Alfv\'en velocity. Recalling that $m/M\,\gg\, 1$, then $\rho_0\,=\,M\, n_0$. The Hawking temperature becomes
\begin{equation}
T_H\approx 2.66~\mbox{K}~\left(\frac{m_p}{M}\right)^{1/2}\left(\frac{1~\mbox{mm}}{R}\right)\left(\frac{1~\mbox{cm}^{-3}}{n_0}\right)^{1/2}\left(\frac{B_0}{\mbox{gauss}}\right)\, ,
\end{equation}
where $m_p$ is the proton mass.

It is possible to imagine configurations where the mix of a magnetic field and an appropriate number density can produce a Hawking temperature larger than that associated to a neutral fluid. Keeping in mind that the size of the nozzle must follows the geometric limit imposed for our approximation.

With the following estimations: the ion mass as the proton mass and the radius of the nozzle considered as $1\mbox{mm}$, then for the number density equal to $n_0=10^{12}$ cm$^{-3}$ and with a magnetic field $B_0\,=\,10^4$ gauss, we obtain $T_H\,\approx\, 0.027~\mbox{K}$. However, the Hawking temperature can grow higher with  other values for larger magnetic field and for hotter plasmas.

\section{Conclusions}

Here we have studied and concluded that a plasma dynamics under the magnetohydrodynamics approximation mimics closely certain aspects of scalar fields in curved space-time just as a normal fluid does. The relative velocity associated to the magnetoacoustic modes and that of the flow itself, generates an analogue horizon equivalent to the event horizon of a Schwarzschild black hole. The similarity between these two physical systems is robust. The Hawking temperature generated by the event horizon it is replicated in this plasma evolving on a background magnetic field chosen adequately. The Alfv\'en velocity  increases the magnetoacoustic speed as shown in \eqref{velocidfluidAlfven}, and since this velocity depends on the external magnetic induction $B_0$, it is feasible to raise it. The density of the plasma is another parameter that affects the Hawking temperature in this plasma. With an increasing density, the hawking temperature decreases. 

Due to the the contribution of the magnetic field then, the equivalent of the Hawking radiation in a plasma becomes larger than that in neutral fluids, and it can even grow higher for larger magnetic fields and temperatures. However, the magnetic induction cannot grow arbitrarily large since there are geometric constrains as pointed in the text. Also the plasma requires temperature as large as $10^4$ to $10^5$ degrees. So the ratio between the temperatures of the the thermal emission and the background decreases moderately, with a factor of $10^6$ or $10^7$. Nevertheless in neutral fluids the ratio is larger, around $10^9$.

In this work we have focused ourselves only in the general setting of the problem. We have not given a specifics of such a system. However, the most recent published results (considering other fluids) give a feeling that this effect is robust and remains detectable directly or indirectly even though the external conditions are not the ideal, see for instance  S. Weinfurtner et al. \cite{unruh3} where the thermal nature of the emission appears to be confirmed and also in the work of Belgiorno et al. \cite{belgiorno}, where the radiation detected, according to the authors, can only be attributed to an equivalent of the Hawking radiation. There are more experiments proposed, for instance in Ref.~\cite{philbin,horstmann}. All these results make feasible that the effect displayed here can be possible measured.

All these elements make this problem more interesting, worth to be studied further. Also, the simplicity and ingenuity displayed in the experiments, contributes strongly in this direction. 




\begin{thebibliography}{17}

\bibitem{unruh1} W. G. Unruh, Phys. Rev. Lett. {\bf 46}, 1351 (1981).

\bibitem{hawking} S. W. Hawking, Nature, (London) {\bf 248}, 30(1974).

\bibitem{unruh2} W. G. Unruh, Phys. Rev. D {\bf 51}, 2827 (1995).

\bibitem{gravit} R. Sch\"{u}tzhold and W. G. Unruh, Phys. Rev. D {\bf 66}, 044019 (2002)

\bibitem{dielec} R. Sch\"{u}tzhold G. Plunien and G. Soff, Phys. Rev. Lett. {\bf 88}, 061101 (2002).

\bibitem{visser} M. Visser, C. Barcel\'o and S. Liberati, Gen. Rel. Grav. {\bf 34}, 1719 (2002).

\bibitem{giova} S. Giovanni, Phys. Rev. Lett. {\bf 94}, 061302 (2005).

\bibitem{garay} L. J. Garay {\it et al.}, Phys. Rev. Lett. {\bf 85}, 4643 (2000).

\bibitem{garay2} L. J. Garay, Int. Jour. Theor. Phys. {\bf 41}, 2073 (2002).

\bibitem{fagno2} S. Fagnocchi, Jour. Phys.: Conference Series {\bf 222}, 012036  (2010).

\bibitem{ralf} R. Sch\"{u}tzhold and W. G. Unruh, Phys. Rev. Lett. {\bf 95}, 031301 (2005).

\bibitem{philbin} T. G. Philbin {\it et al.}, Science {\bf 319}, 1367 (2008).

\bibitem{belgiorno} F. Belgiorno {\it et al.}, Phys. Rev. Lett. {\bf 105}, 203901 (2010).

\bibitem{comments} R. Sch\"{u}tzhold and W. G. Unruh, arXiv:1012.2686v1 [quant-ph], Reply :F. Belgiorno {\it et al.},arXiv:1012.5062v1.

\bibitem{unruh3} S. Weinfurtner {\it et al.}, Phys. Rev. Lett. {\bf 106}, 021302 (2011).

\bibitem{shercliff} J. A. Shercliff, {\it A textbook of magnetohydrodynamics} (Pergamon Press, 1965).

\bibitem{comm} The pressure is defined as $p=p_e+p_i$, where $p_e$ and $p_i$ are the pressures of the electron and ion fluid respectively.

\bibitem{novello}{\it Artificial Black Holes}, edited by M. Novello, M. Visser and G. Volovik (World Scientific, River Edge, USA, 2002).

\bibitem{chen} F. F. Chen, {\it Introduction to plasma physics and controlled fusion} (Plenum Press, 1990).

\bibitem{priest} E. R. Priest, {\it Solar Magnetohydrodynamics}, (D. Reidel Publishing Company, 1982).

\bibitem{visser1998} M. Visser, Phys. Rev. Lett. {\bf 80}, 3436,(1998).


\bibitem{horstmann} B. Horstmann {\it et al.}, Phys. Rev. Lett. {\bf 104}, 250403, (2010).




\end{thebibliography}
\end{document}